\documentclass[11pt]{iopart}

\usepackage{amssymb,bm} 
\usepackage{cite}
\usepackage{graphicx}
\usepackage[colorlinks,citecolor=blue]{hyperref}
\def\band#1{\scriptscriptstyle(#1)}

\def\openone{\leavevmode\hbox{\small1\kern-3.8pt\normalsize1}}%

\begin{document}

\title[The Geometry of (non)-Abelian adiabatic pumping]{The Geometry of (non)-Abelian adiabatic pumping}

\author{Rapha\"{e}l Leone}

\address{54000 Nancy, France}
\ead{raphael.leone@free.fr}
\begin{abstract}
We give a gauge description of the adiabatic
charge pumping in closed systems, both in Abelian and non-Abelian processes, and by means of asymptotic Wilson loops in a suitable parameter manifold. Our geometric formulation provides new insights into this issue, and a very simple algorithm for numerical computations. Indeed, as we show first, discretized Berry--Wilczek--Zee holonomies are easy to implement. Finally, we study non-Abelian pumping in a solvable four-state model, already used in several contexts, to demonstrate the relevance of our approach.
\end{abstract}

%Uncomment for PACS numbers title message
\pacs{03.65 Vf, 05.60.Gg}
% Keywords required only for MST, PB, PMB, PM, JOA, JOB? 
%\vspace{2pc}
%\noindent{\it Keywords}: Article preparation, IOP journals
% Uncomment for Submitted to journal title message
%\submitto{\JPA}
% Comment out if separate title page not required
\maketitle

\section{Introduction}
Geometry and topology play a central role in modern
physics~\cite{Nakahara}, obviously in advanced areas of field
theories, but also in ``standard'' quantum mechanics. Indeed, since
Berry's seminal paper~\cite{Berry}, an
increasing number of quantum phenomena have been well understood through
the mathematical apparatus of gauge theories~\cite{Bleecker,Preparation}.
Some of the most outstanding examples are certainly to be found in
quantum transport physics: Aharonov--Bohm effect, Thouless
pumping, various quantum Hall effects, etc. The geometrical and
topological features in question arise directly from Hilbert
spaces. They are no more seen as simple vector spaces, but as bundles based
on their own projective space~\cite{A-A} or, if the system is
parameter-dependent, on a \textit{parameter
manifold}~\cite{Berry,Simon}.

The latter construction gives a great insight into physics of
adiabatically driven quantum systems. Here, we only need to look at a single level and the initial bundle can
be considerably reduced. For an $n$-fold degenerate
level $(n\geqslant1)$, a ${\rm U}(n)$ gauge structure naturally
appears over parameter manifolds. The physical information is then
entirely carried by Berry~\cite{Berry} ($n=1$) or
Wilczek--Zee~\cite{WZ} ($n>1$) connections. Indeed, they induce parallel
transports in fibres which are tantamount to the Schr\"odinger equation
and determine the adiabatic wavefunction along parameters
evolutions (up to an overall dynamical phase factor). Above all, oriented
loops in parameter manifolds provide \textit{parallel transport
maps} called \textit{holonomies}. In a standard gauge theoretic
language, once given a gauge fixing, holonomies are represented by ${\rm
U}(n)$ matrices known as \textit{Wilson loops}\footnote{Wilson loops are often defined as the trace of
holonomies. We choose an alternative definition, used in several
contexts: Wilson loops as matrix representations of holonomies.}.
Just as representations of physical observables, these matrices
transform \textit{covariantly} with respect to loops' base points.
Holonomies are \textit{de facto} susceptible to enclose some
``elements of reality''. For example, ongoing research in quantum information tries to use them concretely for
implementation of quantum logic operations~\cite{Ekert,Zanardi}. Besides, the gauge-covariant
curvatures, naturally induced by Berry--Wilczek--Zee
connections, may for their part encode some ``local elements of
reality''.\\

The non-degenerate case ($n=1$) is generic and therefore the most
frequently encountered in physical literature. Notably, ${\rm
U}(1)$ Wilson loops are nothing else but the celebrated
\textit{Berry's phase} factor~\cite{Berry} accumulated by the
wavefunction along a cyclic evolution. Since, in this singular
Abelian case, covariance reduces to invariance, Berry's phase is
potentially measurable. As a matter of fact, it has been directly
observed in a variety of interference
experiments~\cite{Tomita,NMR,NQR,Wernsdorfer,Joseph}. This phase has demonstrated its usefulness
to interpret some concepts or phenomena~\cite{Shapere} such as, for instance, the paradigmatic Aharonov--Bohm effect~\cite{A-B} whose
description in terms of Berry's phase was made by Berry himself. We can also mention the macroscopic polarization in crystalline dielectrics~\cite{King,Resta,GoryoKohmoto} expressed through Berry's phases of Bloch states across entire Brillouin zones~\cite{Zak} (also called Zak's phases), or the pumped charge in Cooper pairs pumps~\cite{Toppari,Pekola,TQCP}.
Moreover, locally, Berry's curvature field may be taken into account to construe
some physical effects. It is the case of the anomalous velocity of Bloch electrons, a correction to the usual term --- coming from the band energy dispersion $\mathcal E(\boldsymbol k)$ --- which manifests itself when the crystal momentum $\boldsymbol k$ is moved by external forces~\cite{ChangNiuPRL,ChangNiuPRB,Sundaram,ReviewNiu}.

There is a close relation between Berry's connection and current operators given by partial derivatives (or gradient) of a Hamiltonian $H$ with respect to parameter(s)~\cite{Chern}. This one is at the root of all the great geometrical and topological properties of systems endowed by such a current observable. Physical realizations are potentially multiple. For example, the current of Bloch electrons is proportional to $\partial_{\boldsymbol k} H$, while the current crossing a phase polarized Cooper pair pump is proportional to $\partial_\varphi H$, where $\varphi$ is the superconducting phase bias of the device. It can be shown that, when we consider a non-degenerate eigenstate of $H$, a current operator of this form splits into a usual \textit{dynamical} term (coming from the energy) and a \textit{geometric} correction which can be interpreted in terms of Berry's curvature (see the example of the anomalous velocity). To the latter contribution is assigned the generic name \textit{adiabatic pumped current}. In the context of the famous Thouless pumping~\cite{Thouless}, averaging the pumped current over the whole parameter manifold (a 2-torus) leads to a quantization of the charge. This stems from the topology of the line bundle over the torus, characterized by an invariant integer called first Chern number~\cite{Nakahara,Simon,Chern}. It is, \textit{mutatis mutandis}, the same topological quantization as the Hall conductance for the integer quantum Hall effect~\cite{TKNN,Kohmoto}. Obviously, that kind of topological invariance is highly interesting for metrological issues~\cite{Keller}.\\ 

A. Joye \textit{et al}~\cite{Joye} studied recently the current pumped in the degenerate non-Abelian case, too. They exhibited the differences between the Abelian and non-Abelian cases in a rigorous and somewhat abstract analytical approach. In this paper, we adopt the geometric counterpart of their work. Our aim is to give a more simple and comprehensive picture of adiabatic pumping, mainly for practical computations and later hypothetic applications (some predictions were already done within the framework of superconducting circuits~\cite{Hekking,Ground}). We reserve \sref{sec:level2} to a review
of the basic gauge principles underlying adiabatic dynamics, and
we will emphasize the central role played by
Wilson loops, especially when discretized for numerical purposes.
Then, \sref{sec:level3} is devoted to the main subject. Considering a general $n$-fold degenerate level, we will express any pumped charge in terms of some Wilson loops' element. Contrary to what Wilson loop may seem, cyclic evolutions of the parameters are not required. The geometric difference between the two cases $n=1$ and $n>1$ will be identified. In the non-Abelian case, our holonomic formulation will allow a simple measure of ``non-commutativity'' of pumping cycles. Finally, in \sref{sec:level4}, we make use of a ``toy
model'' already encountered in non-Abelian pumping contexts~\cite{Hekking,Joye} to
compare our previous results with simple analytical expressions.

\section{\label{sec:level2}Geometrical structure underlying quantum adiabaticity}

In this section, we briefly review the peculiar geometry of
quantum systems adiabatically driven by classical parameters.
Broadly speaking, such parameters can be degrees of freedom
having their own dynamics (e.g. nuclei positions in the
Born--Oppenheimer approximation), constraints eventually tuned by
experimentalists (e.g. magnetic fields
and voltages in the context of quantum circuits~\cite{Wendin}), time $t$ itself,
etc. They appear as real arguments of the Hamiltonian, inducing a
parameterized spectrum. If there is a total of $m$ parameters
$x^\mu$, it is natural to represent the ``parameter state'' by a
vector $x=(x^1,\dots,x^m)$ in $\mathbb R^m$. Nevertheless, for
example, a periodicity of $H$ in $x^\mu$ can motivate us to
compactify the ``$\mu$-th dimension'' onto $S^1$. Or, regarding some properties,
\textit{defects} may occur in a certain domain of $\mathbb R^m$ inducing ``holes''
in it (e.g. gauge anomalies). As a generic consequence, we are likely to use an
$m$-dimensional (real) smooth \textit{parameter manifold} $M$, not
necessary $\mathbb R^m$. Setting $\mathcal H$ as the Hilbert space,
we thereby construct a map $H: M\to\mathcal L(\mathcal H)$
assigning to $p$ the Hamiltonian $H(p)$. An evolution of the
parameters (i.e. a path in $M$) induces an evolution of the
wavefunction $\Psi$ (i.e. a path in $\mathcal H$) via the
Schr\"odinger equation. In the well-known adiabatic assumption, under some hypothesis,
the latter obeys to a simple parallel transport rule in an
``eigenspace bundle'', as reviewed below.

\subsection{General settings}

Let us assume that the gap hypothesis is fulfilled for an isolated
eigenvalue $\mathcal E$ of $H$, smoothly defined everywhere on $M$. That is
to say, $\mathcal E$ is a smooth function on $M$ and its finite
multiplicity $n$ is an invariant. Let $\mathfrak E_p\simeq\mathbb
C^n$ be the eigenspace corresponding to $\mathcal E(p)$ for each point $p$
in $M$. In the adiabatic limit, if the state is initially found in
$\mathfrak E_{\gamma(0)}$, it is
well-known~\cite{Kato,Messiah} that along a path $\gamma:\tau\mapsto\gamma(\tau)$, the
wavefunction ``lives'' at each time $\tau$ in $\mathfrak
E_{\gamma(\tau)}$. The whole Hilbert space being useless, one
reduces the structure when ``gluing'' together all the spaces
$\mathfrak E_p$
to yield a smooth vector bundle $\pi:\mathfrak E\to M$ 
with fibres $\pi^{-1}(p)=\mathfrak E_p$. In this bundle point of
view, the adiabatic wavefunction
$\Psi(\tau)\in\pi^{-1}[\gamma(\tau)]$ moves horizontally to
$\gamma$. More precisely, up to an overall phase factor, $\Psi$
obeys to the BWZ parallel transport rule built
as follows. Let $\{|u_a(p)\rangle\in\pi^{-1}(p)\}_{a=1,\dots,n}$
be a local smooth choice of \textit{orthonormal} frames. Within
this \textit{gauge fixing}, the BWZ connection $\omega$
is locally represented by a $\mathfrak{u}(n)$-valued 1-form field
$\mathcal A$: using a coordinate chart $x=(x^1,\dots,x^m)=x^\mu e_\mu$,
$\mathcal A$ reads $A_\mu\,{\rm d}x^\mu$ with $A_{\mu\,b}^a:=\langle
u_a|\partial_\mu u_b\rangle$. 

We now consider a path $\gamma:\tau\mapsto\gamma(\tau)$ in $M$. In
the adiabatic limit, the wavefunction is
$\Psi(\tau)=e^{i\,\eta(\tau)}\,\psi(\tau)$, where
\begin{equation*}
\eta(\tau)=-\frac{1}{\hbar}\int_0^\tau \mathcal E[\gamma(\tau')]\,{\rm d}\tau'
\end{equation*}
is the dynamical phase, and $\psi$ the parallel transport of
$\Psi(0)=\psi(0)$ along $\gamma$. In components, if
$|\psi\rangle:=\psi^a|u_a\rangle$,
\begin{figure}
\centering
\includegraphics{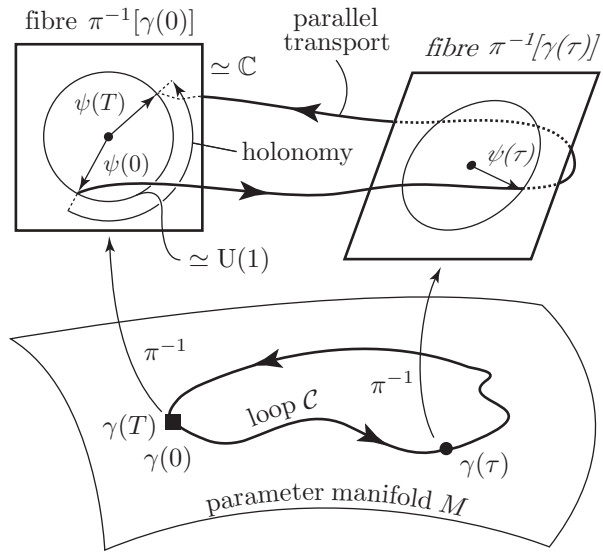}
\caption{Illustration of the gauge structure over a parameter manifold $M$, in the Abelian case $n=1$. Parallel transport in fibres along a loop $\mathcal C\equiv\gamma_{T,0}$ produces a holonomy sending $\psi(0)$ to $\psi(T)$. }\label{figure1}
\end{figure}
the parallel transport equation is
\begin{equation*}
\dot\psi^a+A_{\mu\,b}^{a}\,\dot
x^\mu\,\psi^b=0.
\end{equation*}
Its formal solution is thus
\begin{equation*}
\psi^a(\tau)=W^a_{\;\;b}\big[\gamma_{\tau,0}\big]\,\psi^b(0),
\end{equation*}
where $\gamma_{\tau,0}\equiv\gamma|_{[0,\tau]}$ is the restriction
of $\gamma$ to $[0,\tau]$, and $W$ is the Wilson operator acting
on any path $\alpha$ through
\begin{equation}
W\big[\alpha\big]:=\mathcal P\,e^{-\int_{\alpha}\mathcal A}\in{\rm
U}(n).\label{Wilson}
\end{equation}
$W\big[\alpha\big]$ is commonly called a \textit{Wilson line}, or,
if $\alpha$ is a (based) loop, a \textit{Wilson loop}~\cite{Wilson}.
Finally, one obtains the standard expression of the adiabatic
wavefunction~\cite{WZ}:
\begin{equation}
|\Psi(\tau)\rangle=e^{i\eta(\tau)}W^a_{\;\;b}\big[\gamma_{\tau,0}\big]\,\psi^b(0)\,|u_a[p(\tau)]\rangle.\label{adiabaticwave}
\end{equation}

Let us recall how $\mathcal A$ and $W$ behave under a gauge
transformation (i.e. a change of reference frames) generated by
the smooth map $g:p\mapsto g(p)\in {\rm U}(n)$:
\begin{equation}
\label{gaugetransform}|u_a(p)\rangle\mapsto|\tilde{u}_a(p)\rangle=|u_b(p)\rangle\,g^b_{\;\;a}(p).
\end{equation}
Each component $A_\mu$ transforms under \eref{gaugetransform} in
accordance with the \textit{compatibility condition}:
\begin{equation}
A_\mu(p)\mapsto\widetilde A_\mu(p)=
g^{-1}(p)A_\mu(p)g(p)+g^{-1}(p)\,\partial_\mu g(p),\label{compat}
\end{equation}
i.e. $\mathcal A\mapsto g^{-1}\mathcal A\,g+g^{-1}dg$. As for the
Wilson operator, using the equality
\begin{equation*}
A_\mu[\gamma(\tau)]\,\dot
x^\mu(\tau)=W\big[\gamma_{\tau,0}\big]\,\partial_\tau
W^{-1}\big[\gamma_{\tau,0}\big],
\end{equation*}
it is easy to check that $W\big[\gamma_{\tau,0}\big]$ transforms
according to the \textit{similarity rule}:
\begin{equation}
\label{similarity}W\big[\gamma_{\tau,0}\big]\mapsto \widetilde
W\big[\gamma_{\tau,0}\big]=g^{-1}\big[\gamma(\tau)\big]W\big[\gamma_{\tau,0}\big]g\big[\gamma(0)\big].
\end{equation}
Notably, if $\gamma_{T,0}$ is a loop $\mathcal C$, i.e. if
$\gamma(T)=\gamma(0)=:p_0$, $W\big[\mathcal C\big]$ transforms
covariantly with respect to its base point $p_0$ :
\begin{equation*}
W\big[\mathcal C\big]\mapsto \widetilde W\big[\mathcal
C\big]=g^{-1}(p_0)W\big[\mathcal C\big]g(p_0).
\end{equation*}
Properly, in the basis $\{|u_a(p_0)\rangle\}$, $W\big[\mathcal
C\big]$ represents the parallel transport map
along $\mathcal C$, itself an element of 
the \textit{BWZ holonomy group} based at $p_0$. In particular, when
$n=1$, $W\big[\mathcal C\big]$ reduces to
Berry's phase factor which actually depends only on the unbased loop corresponding to $\mathcal C$ (see figure~\ref{figure1}).
Although it is abusive, by analogy with the Abelian case,
$W\big[\mathcal C\big]$ is often called a non-Abelian geometric phase
when $n>1$. Of course, physical relevance is assigned to its eigenvalues which are gauge invariants of the parallel transport map. A good strategy consists in expressing measurable quantities as their functions (see~\cite{Zhou} in the framework of quantum pumping). In the next section, however, it will be possible to link the pumped current to a single diagonal element of some Wilson loops.

We have not yet spoken about the BWZ curvature $\Omega$. Within a gauge fixing, it is
locally represented by the 2-form field
$\mathcal F:=d\mathcal A+\mathcal A\wedge\mathcal
A=\frac{1}{2}\,F_{\mu\nu}\,{\rm d}x^\mu\wedge{\rm d}x^\nu$, with
\begin{equation*}
F_{\mu\nu}=\partial_\mu A_\nu-\partial_\nu A_\mu+[A_\mu,A_\nu]
\end{equation*}
the antisymmetric curvature tensor. It is obviously gauge
covariant:
\begin{equation*}
F_{\mu\nu}\mapsto\widetilde F_{\mu\nu}=g^{-1}F_{\mu\nu}\,g={\rm
Ad}_{g^{-1}}\big(F_{\mu\nu}\big),
\end{equation*}
i.e. $\mathcal F\mapsto g^{-1}\mathcal F g$. 

\subsection{(Discrete) Wilson loops as fundamental ingredients of
the theory}\label{algorithm}

Because curvature fields and holonomies behave like observables,
both are susceptible to carry some ``elements of reality'', as we
shall see in a while, within the framework of the pumping process. However, we
will regard the latter as the most fundamental objects for two reasons.
\begin{itemize}
\item[(i)] Any component of the curvature tensor can be expressed by
means of an infinitesimal Wilson loop. Indeed, in coordinates,
consider a small rectangular loop $c$ based at $x$ and oriented along the $\mu$-th then the
$\nu$-th directions. Call $\ell_\mu$ its length in the $\mu$-th direction and
$\ell_\nu$ in the $\nu$-th one, with $\ell_\mu\sim\ell_\nu$ (see
\fref{cmunu}). In the appendix, we derive the
following relation:
\begin{equation}
F_{\mu\nu}(x)=\frac{1}{\ell_\mu\ell_\nu}\,\Big\{\openone-W\big[c\big]\Big\}+\Or(\ell),\label{wilsoncurv1}
\end{equation}
where the $\Or$ symbol must be here understood in a matrix norm
sense, and $\ell\sim\ell_\mu,\ell_\nu$ stands for the order of the ``plaquette'' size. The above formula can be
reduced from the anti-Hermitian nature of $F_{\mu\nu}$:
\begin{equation}
F_{\mu\nu}(x)=-\frac{i}{\ell_\mu\ell_\nu}\,\Im\Big\{W\big[c\big]\Big\}+\Or(\ell).\label{wilsoncurv2}
\end{equation}
\begin{figure}
\centering
\includegraphics{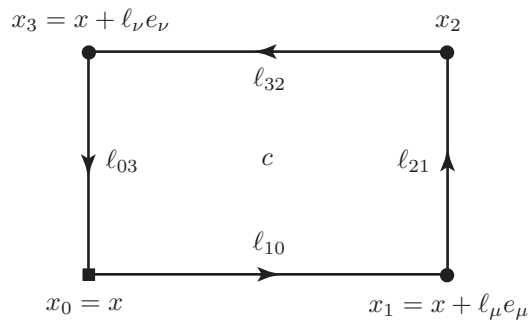}
\caption{The value of the curvature component $F_{\mu\nu}(x)$ is
contained in Wilson loops of small oriented rectangles
$c$, as defined in text. The proof given in the appendix needs a decomposition into four oriented
segments $\ell_{kj}$.}\label{cmunu}
\end{figure}
\item[(ii)] A numerical scheme can easily be implemented to compute
Wilson loops arising in our context. First, note that the naive discretization of formula \eref{Wilson}:
\begin{equation*}
W\big[\mathcal C\big]=\lim_{N\to\infty}\mathcal
P\prod_{j=1}^{N}\Big(\openone-A_\mu(p_j)\Delta x^\mu_j\Big),
\end{equation*}
with $p_0=p_N$ and $\Delta
x^\mu_j:=x^\mu_{j}-x^\mu_{j-1}$, seems inapplicable since it underlies
smoothness in the gauge fixing, while numerical diagonalization
routines ``choose'' eigenvectors in a somewhat erratic manner.
This difficulty is overcome by the contextual definition of
$A_\mu$, leading to
\begin{equation*}
\Big[\openone-A_\mu(p_j)\Delta x^\mu_j\Big]^a_{\;\;b}=\langle
u_a(p_j)|u_b(p_{j-1})\rangle+\Or\big(N^{-2}\big).
\end{equation*}
After introducing the overlap matrices $S(p,q)$, whose elements are
$S^a_{\;\;b}(p,q):=\langle u_a(p)|u_b(q)\rangle$, one obtains a
formulation of great value for numerical treatments: the \textit{
discrete Wilson loop}
\begin{equation}
W\big[\mathcal C\big]=\lim_{N\to\infty}\mathcal P\prod_{j=1}^{N}
S(p_j,p_{j-1}).\label{discretephase}
\end{equation}
\end{itemize}
In the special case $n=1$, the above formula reduces to the
discrete counterpart of Berry's phase factor: it can be seen as asymptotic versions of the
Pancharatnam phase factor~\cite{Pancharatnam} or the Bargmann invariant~\cite{Bargmann}. Covariance of \eref{discretephase} is easily checked. Furthermore, it
allows erratic frame choices along the loop: they cancel out in
pairs, whereas the geometric information is solely extracted. In fact, one has removed the rigid constraints of
differentiability and continuity in frame choices, properties which \textit{in
fine} reveal themselves unnecessary. Building an efficient algorithm based on formula \eref{discretephase} is a simple task: it needs basically scalar and matrix product evaluations, and a diagonalization routine for the Hamiltonian~\cite{Numerical}. For $n=1$, it has been already implemented in
miscellaneous contexts~\cite{King,Mukunda,Resta,Wang,TQCP}. For $n>1$, the result depends on the basis $\{|u_a(p_0)\rangle\}$ chosen on input. Then, in both cases, we let the program ``choose'' the frame $\{|u_a(p)\rangle\}$ of $H(p)$ along the loop. We mainly apply
formula \eref{discretephase} to compute
$W\big[c\big]$ in \eref{wilsoncurv2} as
\begin{equation}
W\big[c\big]=\mathcal P\prod_{j=1}^4
S(x_j,x_{j-1})+\Or(\ell),\label{riquiqui}
\end{equation}
where $x_0=x_4=x$, $x_1$, $x_2$ and $x_3$ are
$c$'s vertices (see \fref{cmunu}). 

\section{\label{sec:level3}Adiabatic charge pumping: a geometric viewpoint}

In this section, we exhibit the geometrical significance of
adiabatic charge pumping. Such a process needs a
quantum system equipped with (i) a current operator $\mathcal I$;
(ii) a Hamiltonian dependence in a controllable ``pumping
parameter'' $\lambda\in I$, where $I$ is an interval of $\mathbb
R$ and (iii) a suitable Hamiltonian gauge such that $\mathcal I$
reads $\frac{q}{\hbar}\,\partial_\lambda H$~\cite{Chern}, where
$q$ is a unit of charge. Physics of pumping becomes richer in the presence of other
parameters. That is why we will assume $H$ defined over an
$(m+1)$-dimensional parameter manifold $M\times I$.
With the current operator is naturally
associated a mean transferred charge $Q$ through the relation $\dot
Q=\langle\mathcal I\rangle$, that is (in units of $q$)
\begin{equation*}
\dot Q(\tau):=\frac{1}{\hbar}\,\langle\Psi(\tau)|\partial_\lambda
H[p(\tau);\lambda]|\Psi(\tau)\rangle.
\end{equation*}

\subsection{Analytic derivation of the pumped charge}

For our purpose, we keep the pumping parameter $\lambda$ constant
throughout quantum evolutions. Once given a path $\gamma$ in $M$,
we define the set $\{\gamma^{\band{\lambda}}\}$ of paths living in
``horizontal'' subregions $\lambda={\rm cst}$ by:
$\gamma^{\band{\lambda}}(\tau):=(\gamma(\tau);\lambda)$. Along
these, the wavefunction $\Psi$ is well defined as a differentiable
function of $\tau$ and $\lambda$. Consequently, we can write
\begin{eqnarray}
\langle\Psi|\partial_\lambda
H|\Psi\rangle&=&\partial_\lambda\langle\Psi|
H|\Psi\rangle-\langle\partial_\lambda\Psi|
H|\Psi\rangle-\langle\Psi|
H|\partial_\lambda\Psi\rangle\nonumber\\
&=&\partial_\lambda \langle
H\rangle-2\hbar\,\Im\big\{\langle\partial_\tau\Psi|\partial_\lambda\Psi\rangle\big\}\,,\label{ham}
\end{eqnarray}
where use has been made of the Schr\"odinger equation. We now
assume an adiabatic evolution for a wavefunction belonging to an
$n$-fold degenerate level. It is convenient to choose an initial
frame such that $\psi^a(0;\lambda)=\delta^{ar}$, with $1\leqslant r\leqslant
n$, and to set
$A_\tau(\tau;\lambda):=A_\mu[\gamma^{\band{\lambda}}(\tau)]\,\dot
x^\mu(\tau)$. Inserting the adiabatic
approximation \eref{adiabaticwave} in \eref{ham} and using
the two relations below:
\begin{eqnarray*}
\partial_{\tau}\! W\big[\gamma^{\band{\lambda}}_{\tau,0}\big]&=&-\,A_\tau(\tau;\lambda)\,W\big[\gamma^{\band{\lambda}}_{\tau,0}\big],\\
\partial_\lambda
W\big[\gamma^{\band{\lambda}}_{\tau,0}\big]&=&-\displaystyle\int_{0}^{\tau}\!W^{-1}\big[\gamma^{\band{\lambda}}_{\tau',0}\big]\,
\partial_\lambda
A_\tau(\tau';\lambda)W\big[\gamma^{\band{\lambda}}_{\tau',0}\big]\,{\rm d}\tau',
\end{eqnarray*}
one obtains, after a little algebra, the instantaneous
\textit{dynamical} and \textit{geometrical}
(or \textit{pumped}) contributions to the
transferred charge. They are respectively
\numparts
\begin{eqnarray}
\label{infdyn}{\rm d} Q_{\rm d}(\tau)&=&\frac{1}{\hbar}\,\partial_\lambda
\mathcal E[\gamma^{\band{\lambda}}(\tau)]\,{\rm d}\tau\qquad{\rm and}\\
\label{infgeo}{\rm d} Q_{\rm
g}^{\band{r}}(\tau)&=&i\Big[W^{-1}\big[\gamma^{\band{\lambda}}_{\tau,0}\big]\,F_{\tau\lambda}(\tau;\lambda)\,W\big[\gamma^{\band{\lambda}}_{\tau,0}\big]\Big]^r_{\;\,r}{\rm d}\tau,
\end{eqnarray}
\endnumparts
with $F_{\tau\lambda}:=\partial_\tau A_\lambda-\partial_\lambda
A_\tau+[A_\tau,A_\lambda]=F_{\mu\lambda}\,\dot x^\mu$.

The dynamical term \eref{infdyn} is obviously gauge invariant, local, and
depends on the time parametrization (i.e. the dynamics of the path). However, it is independent of the initial state in the degenerate subspace. We now
turn our interest to the geometrical contribution \eref{infgeo}. It is clearly sensitive to the initial state ($r$-label) and relies on
an effective gauge structure in two variables $\tau$ and
$\lambda$. We also recognize a component of the so-called
\textit{(effective) twisted curvature tensor} $\mathfrak
F$~\cite{Broda}. Namely,
\begin{equation}
\mathfrak
F_{\tau\lambda}(\tau;\lambda):=W^{-1}\big[\gamma^{\band{\lambda}}_{\tau,0}\big]\,F_{\tau\lambda}(\tau;\lambda)\,W\big[\gamma^{\band{\lambda}}_{\tau,0}\big].\label{twisted}
\end{equation}
Under a gauge transformation $g:(\tau;\lambda)\mapsto
g(\tau;\lambda)$, this one behaves as a Wilson loop based at
$(0;\lambda)$:
\begin{equation*}
\mathfrak F_{\tau\lambda}(\tau;\lambda)\mapsto\widetilde{\mathfrak F}_{\tau\lambda}(\tau;\lambda)=
g^{-1}(0;\lambda)\,\mathfrak
F_{\tau\lambda}(\tau;\lambda)\,g(0;\lambda).
\end{equation*}
But, since one fixed $\psi^a(0;\lambda)=\delta^{ar}$, the initial
gauge freedom is constrained by
$g^{r}_{\;\;a}(0;\lambda)=g^{a}_{\;\;r}(0;\lambda)=\delta_{a,r}$
for each index $a$. That makes the right-hand side of
\eref{infgeo} gauge invariant. Second, anti-Hermiticity of
$F_{\tau\lambda}$ (and so of $\mathfrak F_{\tau\lambda}$)
guarantees the reality of $Q^{\band{r}}_{\rm g}$. Third, its
geometrical nature is due to its invariance under any time
reparameterization $\tau\mapsto\sigma(\tau)$. Indeed, Wilson loops
are themselves insensitive to parametrization, and one has
\begin{equation*}
F_{\sigma\lambda}\,{\rm d}\sigma=F_{\mu\lambda}\,\frac{{\rm d}
x^\mu}{{\rm d}\sigma}\,\frac{{\rm d}\sigma}{{\rm d}\tau}\,{\rm d}\tau=F_{\mu\lambda}\,\frac{{\rm d}
x^\mu}{{\rm d}\tau}\,{\rm d}\tau=F_{\tau\lambda}\,{\rm d}\tau.
\end{equation*}
In the Abelian case, $\mathfrak F\equiv F$ and ${\rm d} Q^{\band{r}}_{\rm g}$
is local, otherwise it is \textit{path dependent}: the
instantaneous pumped charge at any time is related to the whole
history of the wavefunction. We point out that results \eref{infdyn} and \eref{infgeo} are easily generalizable for a vectorial current $\boldsymbol{\mathcal I}=\partial_{\boldsymbol\lambda} H$ with $\boldsymbol\lambda=(\lambda_1,\lambda_2,\dots)$.

\subsection{Geometric interpretation of the pumped charge}In the previous paragraph we showed that the instantaneous pumped
charge between $\tau$ and $\tau+{\rm d}\tau$ is purely geometric and
relies on $\mathfrak F$, a tensor which behaves under a gauge
transformation as a Wilson loop based at $(0;\lambda)$. These
observations motivate us to search for a holonomic interpretation
of $Q_{\rm g}^{\band{r}}$.

\begin{figure}
\centering
\includegraphics{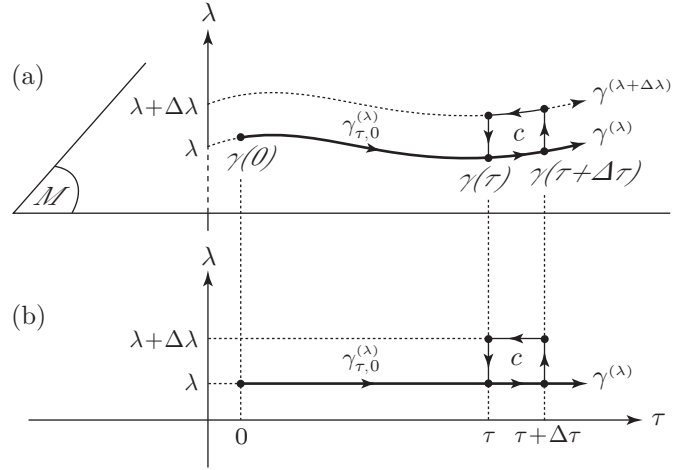}
\caption{(a) The parameter manifold $M$ is symbolized by a
horizontal plane while $\lambda$'s axis is seen in the vertical
direction. The relevant paths $\gamma^{\band{\lambda}}$ are made
at constant $\lambda$. It is shown in the text that the instantaneous
pumped charge can be extracted from a loop
$(\gamma^{\band{\lambda}}_{\tau,0})^{-1}\!\circ
c\circ\gamma^{\band{\lambda}}_{\tau,0}$, where $c$ is a small
rectangular loop oriented along the path and $\lambda$'s axis. (b)
The effective space in variables $\tau$ and $\lambda$.}\label{cheminv}
\end{figure}Let us first have a look at the charge pumped during an arbitrary small time
interval $\Delta\tau$:
\begin{equation*}
\Delta Q^{\band{r}}_{\rm g}(\tau)=i\,\mathfrak
F^{\,r}_{\tau\lambda\;\,r}(\tau;\lambda)\,\Delta\tau+\Or\big[(\Delta\tau)^2\big].
\end{equation*}
We may take advantage of formula \eref{wilsoncurv1} to compute
the component $F_{\tau\lambda}(\tau;\lambda)$ in
\eref{twisted}. If we denote by $c$ the small rectangular
loop pictured in \fref{cheminv}, based at $(\tau;\lambda)$ with
$\Delta\tau\sim\Delta\lambda\sim\Delta$, equality \eref{twisted}
becomes:
\begin{eqnarray*}
\mathfrak
F_{\tau\lambda}(\tau;\lambda)&=&W^{-1}\big[\gamma^{\band{\lambda}}_{\tau,0}\big]\frac{\openone-W\big[c\big]}{\Delta\tau\Delta\lambda}\,W\big[\gamma^{\band{\lambda}}_{\tau,0}\big]+\Or\big(\Delta^3\big)\\
&=&\frac{1}{\Delta\tau\Delta\lambda}\Big\{\openone\!-\!W\big[(\gamma^{\band{\lambda}}_{\tau,0})^{-1}\!\circ
c\circ\gamma^{\band{\lambda}}_{\tau,0}\big]\Big\}\!+\!\Or\big(\Delta^3\big).
\end{eqnarray*}
Setting $\mathcal C$ the loop
$(\gamma^{\band{\lambda}}_{\tau,0})^{-1}\!\circ
c\circ\gamma^{\band{\lambda}}_{\tau,0}$ based at $(0;\lambda)$,
one obtains an alternative formulation of the pumped charge:
\begin{equation}
\Delta Q^{\band{r}}_{\rm
g}(\tau)=\frac{1}{\Delta\lambda}\,\Im\Big\{W^{\,r}_{\;\;r}\big[\mathcal
C\big]\Big\}+\Or\big(\Delta^2\big).\label{localcharge}
\end{equation}
Note that in the Abelian case $W\big[\mathcal C\big]$ is equal to
$W\big[c\big]$ and the rectangle $c$ alone suffices for the computation. That represents geometrically the difference between the Abelian and non-Abelian pumpings, in terms of locality or non locality. Formula \eref{localcharge}, when combined
with \eref{discretephase} and \eref{riquiqui}, is very useful for
numerical purposes, as we will see later with an explicit
example.

\begin{figure}
\centering
\includegraphics{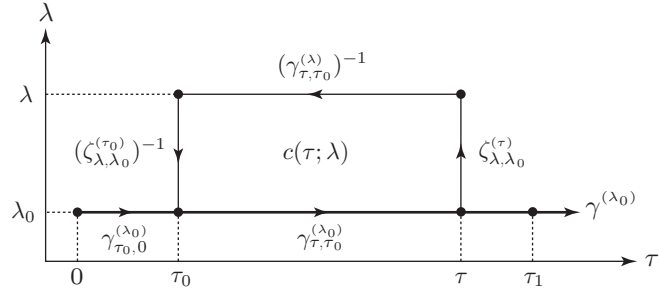}
\caption{An illustration of the various oriented segments and
loops defined in the text to compute the pumped charge between times
$\tau_0$ and $\tau_1$.}\label{long}
\end{figure}

As a second step, we would like to derive a similar expression for
finite $\Delta\tau$. To this end, we will consider the charge
$Q_{\rm g}\big[\gamma^{\band{\lambda_0}}_{\tau_1,\tau_0}\big]$
pumped along $\gamma^{\band{\lambda_0}}$ between any two times
$\tau_0$ and $\tau_1$:
\begin{equation}
Q^{\band{r}}_{\rm
g}\big[\gamma^{\band{\lambda_0}}_{\tau_1,\tau_0}\big]=i\int_{\tau_0}^{\tau_1}\mathfrak
F^{\,r}_{\tau\lambda\;\,r}(\tau;\lambda_0)\,{\rm d}\tau.\label{but}
\end{equation}
Let
$\{\zeta^{\band{\tau}}_{\lambda,\lambda_0}\,|\,(\tau,\lambda)\in[\tau_0,\tau_1]\times
I\}$ be the set of ``vertical'' segments oriented from
$(\tau;\lambda_0)$ to $(\tau;\lambda)$ in the effective space (see \fref{long}).
Then, we define rectangular loops $c(\tau;\lambda)$ as follows:
\begin{equation*}
c(\tau;\lambda):=(\zeta^{\band{\tau_0}}_{\lambda,\lambda_0})^{-1}\circ(\gamma^{\band{\lambda}}_{\tau,\tau_0})^{-1}\circ\zeta^{\band{\tau}}_{\lambda,\lambda_0}\circ\gamma^{\band{\lambda_0}}_{\tau,\tau_0}.
\end{equation*}
Setting
$T(\tau;\lambda):=W\big[\gamma^{\band{\lambda}}_{\tau,\tau_0}\circ\zeta^{\band{\tau_0}}_{\lambda,\lambda_0}\big]$,
the Wilson loop $W\big[c(\tau;\lambda)\big]$ can be expressed as
\begin{equation}
\label{looppath}W\big[c(\tau;\lambda)\big]=T^{-1}(\tau;\lambda)\,W\big[\zeta^{\band{\tau}}_{\lambda,\lambda_0}\big]\,T(\tau;\lambda_0),
\end{equation}
where we have noticed the property
\begin{equation}
T(\tau;\lambda_0)=W\big[\gamma^{\band{\lambda_0}}_{\tau,\tau_0}\big].\label{tt}
\end{equation}
Thanks to the similarity rule \eref{similarity},
$W\big[c(\tau;\lambda)\big]$ in equation \eref{looppath} appears to be
the image of $W\big[\zeta^{\band{\tau}}_{\lambda,\lambda_0}\big]$
under the gauge transformation generated by $T$~\cite{Karp}. As for
$A_{\lambda}$, it transforms according to the compatibility
condition \eref{compat}:
\begin{equation*}
A_\lambda\mapsto \widetilde{A}_\lambda=T^{-1}A_\lambda\,
T+T^{-1}\partial_\lambda T.
\end{equation*}
It is straightforward to verify that the partial derivative of
$\widetilde A_\lambda$ with respect to $\tau$ is merely equal to $\widetilde
F_{\tau\lambda}=T^{-1}F_{\tau\lambda}T$. Or, in the integral form:
\begin{equation}
\widetilde A_\lambda(\tau;\lambda)=\int_{\tau_0}^\tau
T^{-1}(\tau';\lambda)F_{\tau\lambda}(\tau';\lambda)T(\tau';\lambda)\,{\rm d}\tau'.\label{integral}
\end{equation}
The last result allows a curvature formulation of the Wilson loop
$W\big[c(\tau;\lambda)\big]$, through a ``$\lambda$-ordered''
surface integral:
\begin{eqnarray*}
W\big[c(\tau;\lambda)\big]&=&\widetilde
W\big[\zeta^{\band{\tau}}_{\lambda,\lambda_0}\big]=\mathcal
P_\lambda\,
e^{-\int_{\lambda_0}^\lambda\widetilde{A}_\lambda(\tau;\lambda')\,{\rm d}\lambda'}\nonumber\\
&=&\mathcal
P_\lambda\,\exp\Bigg\{\!\!-\frac{1}{2}\,\int_{\Sigma(\tau;\lambda)}\!\!\!\!\!
T^{-1}F_{\tau\lambda}T\;{\rm d}\tau\wedge {\rm d}\lambda\Bigg\},
\end{eqnarray*}
where $\Sigma(\tau;\lambda)$ is the well-oriented rectangular surface suspending the loop $c(\tau;\lambda)$ in the
effective space. This is
\textit{en passant} an apparition of the so-called non-Abelian
Stokes theorem~\cite{Karp,Broda}. The connection between the pumped
charge \eref{but} and our loops is established when taking
$c(\tau_1;\lambda)$ for a small difference
$\Delta\lambda:=\lambda-\lambda_0$:
\begin{equation*}
W\big[c(\tau_1;\lambda)\big]=\mathcal P_\lambda\,
e^{-\int_{\lambda_0}^{\lambda_0+\Delta\lambda}\widetilde{A}_\lambda(\tau_1;\lambda')\,{\rm d}\lambda'}
=\openone-\widetilde
A_\lambda(\tau_1;\lambda_0)\,\Delta\lambda+\Or\big[(\Delta\lambda)^2\big].
\end{equation*}
Using equation \eref{integral} and property \eref{tt}, one obtains
\begin{equation}
W\big[c(\tau_1;\lambda)\big]=\openone\!-\!\Delta\lambda\!\!\int_{\tau_0}^{\tau_1}\!\!\!W^{-1}\big[\gamma^{\band{\lambda_0}}_{\tau,\tau_0}\big]
F_{\tau\lambda}(\tau;\lambda_0)W\big[\gamma^{\band{\lambda_0}}_{\tau,\tau_0}\big]{\rm d}\tau+\Or\big[(\Delta\lambda)^2\big].\label{holol}
\end{equation}
We set $\mathcal C=(\gamma^{\band{\lambda}}_{\tau_0,0})^{-1}\circ
c(\tau_1;\lambda)\circ\gamma^{\band{\lambda}}_{\tau_0,0}$, whose
corresponding Wilson loop, deduced from \eref{holol}
and \eref{twisted}, is
\begin{equation}
W\big[\mathcal
C\big]=\openone-\Delta\lambda\!\int_{\tau_0}^{\tau_1}\!\!\mathfrak
F_{\tau\lambda}(\tau;\lambda_0)\,{\rm d}\tau+\Or\big[(\Delta\lambda)^2\big].\label{presque}
\end{equation}
After comparing \eref{presque} with \eref{but} and using again
anti-Hermiticity of $\mathfrak F_{\tau\lambda}$, one obtains the
final expression:
\begin{equation}
\label{deltageo}Q^{\band{r}}_{\rm
g}\big[\gamma_{\tau_1,\tau_0}^{\band{\lambda_0}}\big]=\frac{1}{\Delta\lambda}\,\Im\Big\{W^{\,r}_{\;\;r}\big[\mathcal
C\big]\Big\}+\Or(\Delta\lambda).
\end{equation}

We achieved our purpose: computations of pumped charges need in
any case a simple Wilson loop evaluation by way of
relation \eref{discretephase}. Let us conclude this section with
a consideration on loops in parameter manifold: non-Abelian charge pumping will give rise to non-commuting loops in
the sense that loops $\Gamma_1^{\band{\lambda}}$
and $\Gamma_2^{\band{\lambda}}$ based at the same point will generically be such that
$Q^{\band{r}}_{\rm
g}\big[\Gamma_1^{\band{\lambda}}\circ\Gamma_2^{\band{\lambda}}\big]\neq
Q^{\band{r}}_{\rm
g}\big[\Gamma_2^{\band{\lambda}}\circ\Gamma_1^{\band{\lambda}}\big]$.
If $Q^{\band{r}}_{\rm g}\big[\Gamma_1^{\band{\lambda}}\big]$ and
$Q^{\band{r}}_{\rm g}\big[\Gamma_2^{\band{\lambda}}\big]$ rely
respectively on loops $\mathcal C_1$ and $\mathcal C_2$ in
equation \eref{deltageo}, one can measure their ``non-commutativity''
through
\begin{equation*}
Q^{\band{r}}_{\rm
g}\big[\Gamma_1^{\band{\lambda}}\circ\Gamma_2^{\band{\lambda}}\big]-Q^{\band{r}}_{\rm
g}\big[\Gamma_2^{\band{\lambda}}\circ\Gamma_1^{\band{\lambda}}\big]=
\frac{1}{\Delta\lambda}\,\Im\Bigg\{\Big[W\big[\mathcal
C_1\big],W\big[\mathcal C_2\big]\Big]^r_{\;\,r}\Bigg\}+\Or(\Delta\lambda).
\end{equation*}
Some experiments are already proposed to demonstrate the non-commutativity of pumping cycles~\cite{Zhou,Hekking}.

\section{\label{sec:level4}An illustrative example}

\begin{figure}
\centering
\includegraphics{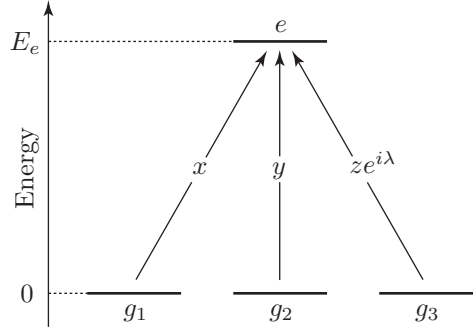}
\caption{Relative couplings between the excited state $|e\rangle$ and the three ground states $|g_1\rangle$, $|g_2\rangle$ and $|g_3\rangle$, in the four-state model discussed in \sref{sec:level4}.}\label{couplages}
\end{figure}

We illustrate the results derived in the previous section with a
well-known model appearing in atomic~\cite{Zanardi} and
superconducting~\cite{Hekking} systems. It consists of an
excited state $|e\rangle$ parametrically coupled to three
independent ground states
$|g_1\rangle$, $|g_2\rangle$ and $|g_3\rangle$. In~\cite{Hekking} the
Hamiltonian reads
\begin{equation*}
H(\boldsymbol r;\lambda)=E_e|e\rangle\langle
e|+x\,|e\rangle\langle g_1|+y\,|e\rangle\langle
g_2|+z\,e^{i\lambda}\,|e\rangle\langle g_3|+{\rm
h.c}.,
\end{equation*}
where $x$, $y$ and $z$ are three real parameters, $\boldsymbol
r$ is the vector $(x,y,z)$ in the parameter space $\mathbb R^3$,
and $E_e$ is the energy assigned to state $|e\rangle$ (eventually
zero), while energies of the ground states are set to zero (see \fref{couplages}). In the
context of~\cite{Hekking}, $H(\boldsymbol r;\lambda)$ is a
truncated Hamiltonian in the Coulomb blockade regime where
$|e\rangle$, $|g_1\rangle$, $|g_2\rangle$ and $|g_3\rangle$ are the four
relevant charge states (i.e. Cooper pairs in excess on different
coupled islands). The parameters are then Josephson couplings
between islands, the pumping parameter $\lambda$ is the phase bias
between two superconducting leads and the current operator is the
Cooper pair current from one lead to the other. Experimentally
speaking, the parameters $x$, $y$ and $z$ live only in a
restricted range of values. One may ask $\boldsymbol r$ to belong
to a connected subset of $\mathbb R^3$ but this restriction is
somewhat irrelevant. When $\boldsymbol r\neq\boldsymbol 0$, the Hamiltonian has three
distinct eigenvalues: $E_-$, $E_0\equiv0$ and $E_+$, the second
being doubly-degenerate over $\mathbb R^3\backslash\{\boldsymbol 0\}$ while the non-degenerate
eigenvalues $E_\pm$ depend only on $r:=\|\boldsymbol
r\|=\sqrt{x^2+y^2+z^2}$:
\begin{eqnarray*}
E_\pm(r)=\frac{1}{2}\Big(E_e\pm\sqrt{E_e^2+4\,r^2}\Big).
\end{eqnarray*}
In the special case $\boldsymbol r=\boldsymbol 0$, the level $E_0$
becomes at least three-fold degenerate: the origin is thus a
defect for the ${\rm U}(2)$ gauge structure and we reduce the
parameter space to $\mathbb{R}^{3}\backslash\{\boldsymbol 0\}$. In order to perform non-Abelian pumping
we focus on the eigenspaces associated with $E_0$. A smooth
choice of orthonormal frame $\{|u_1\rangle,|u_2\rangle\}$ can be
made over $\mathbb R^3\backslash(Oz)\times I$ as follows:
\numparts
\begin{eqnarray}
\label{vec1}|u_1(\boldsymbol
r;\lambda)\rangle&=&\frac{y\,|g_1\rangle-x\,|g_2\rangle}{\sqrt{x^2+y^2}}\\
\label{vec2}|u_2(\boldsymbol
r;\lambda)\rangle&=&\frac{-ze^{i\lambda}\big(x\,|g_1\rangle+y\,|g_2\rangle\big)+\big(x^2+y^2\big)|g_3\rangle}{\sqrt{(x^2+y^2)(x^2+y^2+z^2)}}\,.
\end{eqnarray}
\endnumparts
A similar and compatible choice can also be found over e.g.
$\mathbb R^3\backslash(Oy)\times I$ to cover entirely the
parameter space:
\begin{eqnarray*}
|\tilde{u}_1(\boldsymbol
r;\lambda)\rangle&=&\frac{ze^{i\lambda}|g_1\rangle-x\,|g_3\rangle}{\sqrt{x^2+z^2}}\\
|\tilde{u}_2(\boldsymbol
r;\lambda)\rangle&=&\frac{-y\big(x\,|g_1\rangle+ze^{-i\lambda}|g_3\rangle\big)+\big(x^2+z^2\big)|g_2\rangle}{\sqrt{(x^2+z^2)(x^2+y^2+z^2)}}\,.
\end{eqnarray*}
We note that all these states are invariant under a rescaling
$\boldsymbol r\mapsto k\,\boldsymbol r$, $k>0$.
\begin{figure}
\centering
\includegraphics{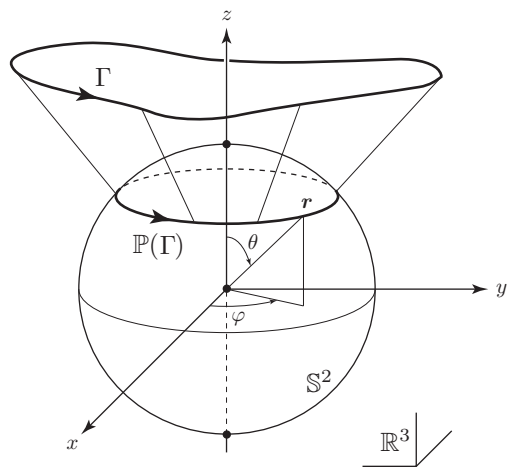}
\caption{The charge pumped along a path $\Gamma$ in $\mathbb R^3$ is equal to the one pumped along $\Gamma$'s radial projection onto the unit sphere $\mathbb S^2$. Here, $\Gamma$ is a loop and its projection $\mathbb P(\Gamma)$ a circle.}\label{figure6}
\end{figure}
Therefore, it is
sufficient to retract $\mathbb{R}^{3}\backslash\{\boldsymbol 0\}$
onto its embedded unit sphere $\mathbb S^2$: the pumped charge
along a path $\gamma$ in $\mathbb R^3$ will be equal to the pumped
charge along its radial projection $\mathbb P(\gamma)$ onto
$\mathbb S^2$ (see figure~\ref{figure6}). Using \textit{local} spherical coordinates on
$\mathbb S^2$: $\boldsymbol
r=(\sin\theta\cos\varphi,\sin\theta\sin\varphi,\cos\theta)$, the
eigenstates \eref{vec1} and \eref{vec2} read:
\begin{eqnarray*}
|u_1(\boldsymbol
r;\lambda)\rangle&=&\sin\varphi\,|g_1\rangle-\cos\varphi\,|g_2\rangle\nonumber\\
|u_2(\boldsymbol
r;\lambda)\rangle&=&-e^{i\lambda}\cos\theta\big(\cos\varphi\,|g_1\rangle+\sin\varphi\,|g_2\rangle\big)+\sin\theta\,|g_3\rangle\nonumber.
\end{eqnarray*}
Within this choice, the components of the effective gauge fields
have simple expressions:
\begin{eqnarray*}
A_\tau(\tau;\lambda)&=&-\cos[\theta(\tau)]\,\dot\varphi(\tau)\left(\begin{array}{cc}0&-e^{i\lambda}\\e^{-i\lambda}&0\end{array}\right),\\
A_\lambda(\tau;\lambda)&=&i\cos^2[\theta(\tau)]\left(\begin{array}{cc}0&0\\0&1\end{array}\right),
\end{eqnarray*}
and:
\begin{eqnarray*}
F_{\tau\lambda}(\tau;\lambda)=\frac{\sin[2\theta(\tau)]}{2\,i}\!\left(\begin{array}{cc}0&\sin[\theta(\tau)]\dot\varphi(\tau)e^{i\lambda}\\ \sin[\theta(\tau)]\dot\varphi(\tau)e^{-i\lambda}&2\dot\theta(\tau)\end{array}\right).
\end{eqnarray*}
The simplicity of the above relations allows an exact computation
of the Wilson line involved by the path followed between times 0
and $\tau$:
\begin{equation*}
W\big[\gamma^{\band{\lambda}}_{\tau,0}\big]=\mathcal
P_\tau\exp\Bigg\{-\int_{0}^{\tau}\!\!A_\tau(\tau;\lambda)\,{\rm d}\tau\Bigg\}
=\left(\begin{array}{cc}\cos(\alpha_{\tau,0})&-e^{i\lambda}\sin(\alpha_{\tau,0})\\e^{-i\lambda}\sin(\alpha_{\tau,0})&\cos(\alpha_{\tau,0})\end{array}\right)
\end{equation*}
where we set:
\begin{equation*}
\alpha_{\tau,0}=\int_{0}^{\tau}\cos[\theta(\tau')]\,\dot\varphi(\tau')\,{\rm d}\tau'=\int_{\gamma_{\tau,0}}\!\!\cos\theta\,{\rm d}\varphi.
\end{equation*}
One has thereby the basic ingredients to calculate via
equation \eref{infgeo} the transferred charges $Q^{\band{1}}_{\rm g}$
and $Q^{\band{2}}_{\rm g}$ along $\gamma^{\band{\lambda}}$.
Namely,
\numparts
\begin{eqnarray}
Q_{\rm
g}^{\band{1}}\big[\gamma^{\band{\lambda}}_{\tau,0}\big]&=&\sin^2(\alpha_{\tau,0})\sin^2[\theta(\tau)],\label{un}\\
Q_{\rm
g}^{(2)}\big[\gamma^{\band{\lambda}}_{\tau,0}\big]&=&\cos^2(\alpha_{\tau,0})\sin^2[\theta(\tau)]-\sin^2[\theta(0)]\label{deux}.
\end{eqnarray}
\endnumparts
Note that they do not depend on $\lambda$'s magnitude; here, $\lambda$
is a pure pumping parameter with no observable effects. It is
straightforward to check that, once the vertical Wilson
line
\begin{eqnarray*}
W(\zeta^{\band{\tau}}_{\lambda+\Delta\lambda,\lambda})=\mathcal
P_\lambda\exp\Bigg\{-\int_{\lambda}^{\lambda+\Delta\lambda}\!\!A_\tau(\tau;\lambda')\,{\rm d}\lambda'\Bigg\}=\left(\begin{array}{cc}1&0\\0&e^{-i\cos^2[\theta(\tau)]\Delta\lambda}\end{array}\right)
\end{eqnarray*}
is obtained, formula \eref{deltageo} yields to the same results \eref{un} and \eref{deux}. In particular, if at time $T$ one has
$\boldsymbol r(T)=\boldsymbol r(0)$, the path
$\gamma^{\band{\lambda}}_{T,0}$ is a loop $\Gamma$ and the two
pumped charges above are opposite: $Q_{\rm
g}^{\band{1}}\big[\Gamma\big]=\sin^2(\alpha_{T,0})\sin^2[\theta(0)]=-Q_{\rm
g}^{\band{2}}\big[\Gamma\big]$.
\begin{figure}
\includegraphics{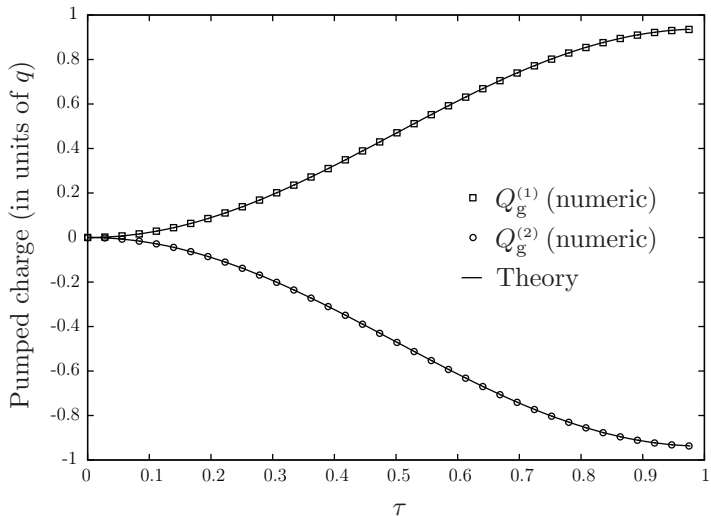}
\centering
\caption{Analytical and numerical plots of the pumped charge in the circular example given in the text, for radius 1, $\cos\theta=1/4$ and $\lambda=0$. The algorithm is based on formulas \eref{discretephase} and \eref{localcharge}.}
\label{transfert}
\end{figure}

We conclude that section by a numerical test in the present non-Abelian example (for an Abelian case, see e.g. \cite{TQCP}). The algorithm is built on the discussion of subsection \ref{algorithm} applied to formula \eref{localcharge}. To check its validity, consider a mere situation: a circle whose axis is $(Oz)$ and such that $\cos\theta=1/4$. Whatever $\lambda$ be, if the circle is followed at constant velocity $\dot\varphi$ from $\tau=0$ to $\tau=1$, expressions \eref{un} and \eref{deux} become:
\begin{equation*}
Q_{\rm g}^{\band{1}}\big[\gamma^{\band{\lambda}}_{\tau,0}\big]=\frac{15}{16}\,\sin^2\bigg(\frac{\pi}{2}\,\tau\bigg)=-Q_{\rm g}^{\band{2}}\big[\gamma^{\band{\lambda}}_{\tau,0}\big].
\end{equation*}
In \fref{transfert}, the above expression together with the numerical result is plotted, and we find that they perfectly coincide.

\section{Conclusion}

We have shown the relevance of Wilson loops to describe
adiabatic charge pumping, mainly for practical purposes. In that pure geometric approach, one may particularly take advantage of
formulas \eref{discretephase} and \eref{localcharge} to compute
pumped charges in any problem. In the future, it should be interesting to investigate the geometry of this process in open systems, and to have a look at topological issues in the non-Abelian case. 

\ack

I am grateful to Bertrand Berche for his helpful suggestions and comments.

\appendix

\section*{Appendix. Curvature extraction from rectangles}\label{holocurv}

In this appendix, we give a proof of formula \eref{wilsoncurv1}.
First of all, we denote by $x_0=x$, $x_1$, $x_2$ and $x_3$ the
vertices of the oriented rectangle $c$, as
depicted in \fref{cmunu}, such that $x_1=x+\ell_\mu\,e_\mu$
and $x_3=x+\ell_\nu\,e_\nu$. Then, we decompose the loop into four
oriented segments joining its vertices: $\ell_{10}$, $\ell_{21}$,
$\ell_{32}$ and $\ell_{03}$, with $\ell_{kj}$ going from $x_j$ to
$x_k$ (see \fref{cmunu}). It suffices to treat $\ell_{10}$ and to apply the result to
the others. This path can be parameterized as follows:
\begin{equation*}
\ell_{10}:[0,1]\ni s\mapsto\ell_{10}(s)=x_0+s(x_1-x_0)=x_0+s\,\ell_\mu\,e_\mu\in\mathbb
R^m.
\end{equation*}
Within a gauge choice, its corresponding Wilson line reads:
\begin{equation*}
W[\ell_{10}]=e^{-\int_0^1A_\sigma[\ell_{10}(s)]\frac{{\rm d}}{{\rm d}s}\big[\ell^\sigma_{10}(s)\big]{\rm d}s}=e^{-\ell_\mu\int_0^1A_\mu[\ell_{10}(s)]{\rm d}s}.
\end{equation*}
Expanding the path-ordered exponential as a usual Dyson series, one
picks the expression of $W[\ell_{10}]$ up to the second order in
$\ell_\mu$ (in a matrix norm sense):
\begin{eqnarray*}
\fl W[\ell_{10}]=\openone+\sum_{k=1}^{\infty}(-\ell_\mu)^k\!\!\int_0^1\!\!{\rm d}s_1\!\int_0^{s_1}\!\!\!{\rm d}s_2\,\cdots\!\int_0^{s_{k-1}}\!\!\!\!{\rm d}s_k
\Big\{A_{\mu}[\ell_{10}(s_1)]A_{\mu}[\ell_{10}(s_2)]\cdots A_{\mu}[\ell_{10}(s_k)]\Big\}\\
\fl\phantom{W[\ell_{10}]}=\openone-\ell_\mu\int_0^1\!{\rm d}s\,A_{\mu}[\ell_{10}(s)]+\ell_\mu^2
\int_0^1\!{\rm d}s\int_0^s\!{\rm d}s'\,A_{\mu}[\ell_{10}(s)]A_{\mu}[\ell_{10}(s')]+\Or\big(\ell_\mu^3\big)\\
\fl\phantom{W[\ell_{10}]}=\openone-A_\mu(x_0)\,\ell_\mu +\frac{1}{2}\,\Big\{A^2_\mu(x_0)-
\partial_\mu A_\mu(x_0)\Big\}\,\ell_\mu^2+\Or\big(\ell_\mu^3\big)\qquad{\rm or}\\
\fl\phantom{W[\ell_{10}]}=\openone-A_\mu(x_1)\,\ell_\mu +\frac{1}{2}\,\Big\{A^2_\mu(x_1)+
\partial_\mu A_\mu(x_1)\Big\}\,\ell_\mu^2+\Or\big(\ell_\mu^3\big).
\end{eqnarray*}
Therefore, the Wilson loop can be expressed as the product
\begin{eqnarray*}
\fl W\big[c\big]=W\big[\ell_{03}\big]\cdot W\big[\ell_{32}\big]\cdot W\big[\ell_{21}\big]\cdot W\big[\ell_{10}\big]\\
\fl\phantom{W\big[c\big]}=\phantom{\cdot}\,\bigg(\openone+A_\nu(x_0)\,\ell_\nu+\frac{1}{2}\,\Big\{A^2_\nu(x_0)+
\partial_\nu A_\nu(x_0)\Big\}\,\ell_\nu^2\bigg)\\
\fl\phantom{W\big[c\big]=}\cdot\bigg(\openone+A_\mu(x_3)\,\ell_\mu+\frac{1}{2}\,\Big\{A^2_\mu(x_3)+
\partial_\mu A_\mu(x_3)\Big\}\,\ell_\mu^2\bigg)\\
\fl\phantom{W\big[c\big]=}\cdot\bigg(\openone-A_\nu(x_1)\,\ell_\nu+\frac{1}{2}\,\Big\{A^2_\nu(x_1)-
\partial_\nu
A_\nu(x_1)\Big\}\,\ell_\nu^2\bigg)\\
\fl\phantom{W\big[c\big]=}\cdot\bigg(\openone-A_\mu(x_0)\,\ell_\mu+\frac{1}{2}\,\Big\{A^2_\mu(x_0)-
\partial_\mu A_\mu(x_0)\Big\}\,\ell_\mu^2\bigg)+\Or\big(\ell^3\big),
\end{eqnarray*}
where $\ell$ stands for the order of the rectangle size:
$\ell\sim\ell_\mu,\ell_\nu$. Finally, using
$A_\nu(x_1)=A_\nu(x_0)+\ell_\nu\partial_\nu A_\nu(x_0)+\Or\big(\ell^2\big)$ and $A_\mu(x_3)=A_\mu(x_0)+\ell_\mu\partial_\mu
A_\mu(x_0)+\Or\big(\ell^2\big)$, one obtains
\begin{eqnarray*}
W\big[c\big]&=&\openone-\Big\{\partial_\mu
A_\nu(x_0)-\partial_\nu
A_\mu(x_0)+\big[A_\mu(x_0),A_\nu(x_0)\big]\Big\}\,\ell_\mu\ell_\nu+\Or\big(\ell^3\big)\\
&=&\openone-F_{\mu\nu}(x)\,\ell_\mu\ell_\nu+\Or\big(\ell^3\big).
\end{eqnarray*}

\section*{References}

\bibliography{pumpingIOP}

\providecommand{\newblock}{}
\begin{thebibliography}{10}
\expandafter\ifx\csname url\endcsname\relax
  \def\url#1{{\tt #1}}\fi
\expandafter\ifx\csname urlprefix\endcsname\relax\def\urlprefix{URL }\fi
\providecommand{\eprint}[2][]{\url{#2}}
% Bibliography created with iopart-num v2.1
% /biblio/bibtex/contrib/iopart-num

\bibitem{Nakahara}
Nakahara M 2003 {\em Geometry, Topology and Physics\/} (London: Taylor and
  Francis)

\bibitem{Berry}
Berry M~V 1984 {\em Proc. R. Soc. A\/} {\bf 392} 45

\bibitem{Bleecker}
Bleecker D 1981 {\em Gauge Theory and Variational Principle\/} (Reading, MA:
  Addison-Wesley)

\bibitem{Preparation}
Svetlichny G 1999  (\textit{Preprint} \eprint{math-ph/9902027})

\bibitem{A-A}
Aharonov Y and Anandan J 1987 {\em Phys. Rev. Lett.\/} {\bf 58} 1593

\bibitem{Simon}
Simon B 1983 {\em Phys. Rev. Lett.\/} {\bf 51} 2167

\bibitem{WZ}
Wilczek F and Zee A 1984 {\em Phys. Rev. Lett.\/} {\bf 52} 2111

\bibitem{Ekert}
Ekert A, Ericsson M, Hayden M, Inamori H, Jones H~A, Oi D~K~L and Vedral V 2000
  {\em J. Mod. Opt.\/} {\bf 47} 2501 (\textit{Preprint}
  \eprint{quant-ph/0004015})

\bibitem{Zanardi}
Recati A, Calarco T, Zanardi P, Cirac J~I and Zoller P 2002 {\em Phys. Rev.
  A\/} {\bf 66} 032309 (\textit{Preprint} \eprint{quant-ph/0204030})

\bibitem{Tomita}
Tomita A and Chiao R~Y 1986 {\em Phys. Rev. Lett.\/} {\bf 57} 937

\bibitem{NMR}
Suter D, Chingas G~C, Harris R~A and Pines A 1987 {\em Mol. Phys.\/} {\bf 61}
  1327

\bibitem{NQR}
Tycko R 1987 {\em Phys. Rev. Lett.\/} {\bf 58} 2281--2284

\bibitem{Wernsdorfer}
Wernsdorfer W and Sessoli R 1999 {\em Science\/} {\bf 284} 133

\bibitem{Joseph}
Leek P~J, Fink J~M, Blais A, Bianchetti R, G\"oppl M, Gambetta J~M, Schuster
  D~I, Frunzio L, Schoelkopf R~J and Wallraff A 2007 {\em Science\/} {\bf 318}
  1889 (\textit{Preprint} \eprint{cond-mat/0711.0218})

\bibitem{Shapere}
Wilczek F and Shapere A 1989 {\em Geometric Phases in Physics\/} (Singapore:
  World Scientific)

\bibitem{A-B}
Aharonov Y and Bohm D 1959 {\em Phys. Rev.\/} {\bf 115} 485

\bibitem{King}
King-Smith R~D and Vanderbilt D 1993 {\em Phys. Rev. B\/} {\bf 47} 1651

\bibitem{Resta}
Resta R 1994 {\em Rev. Mod. Phys.\/} {\bf 66} 899

\bibitem{GoryoKohmoto}
Goryo J and Kohmoto M 2002 {\em Phys. Rev. B\/} {\bf 66} 085118

\bibitem{Zak}
Zak J 1989 {\em Phys. Rev. Lett.\/} {\bf 62} 2747

\bibitem{Toppari}
Aunola M and Toppari J~J 2003 {\em Phys. Rev. B\/} {\bf 68} 020502
  (\textit{Preprint} \eprint{cond-mat/0303176})

\bibitem{Pekola}
M\"ott\"onen M, Vartiainen J~J and Pekola J~P 2008 {\em Phys. Rev. Lett.\/}
  {\bf 100} 177201 (\textit{Preprint} \eprint{cond-mat/0710.5623})

\bibitem{TQCP}
Leone R and L\'evy L 2008 {\em Phys. Rev. B\/} {\bf 77} 064524
  (\textit{Preprint} \eprint{cond-mat/0711.0586})

\bibitem{ChangNiuPRL}
Chang M~C and Niu Q 1995 {\em Phys. Rev. Lett.\/} {\bf 75} 1348
  (\textit{Preprint} \eprint{cond-mat/9505021})

\bibitem{ChangNiuPRB}
Chang M~C and Niu Q 1996 {\em Phys. Rev. B\/} {\bf 53} 7010 (\textit{Preprint}
  \eprint{cond-mat/9511014})

\bibitem{Sundaram}
Sundaram G and Niu Q 1999 {\em Phys. Rev. B\/} {\bf 59} 14915
  (\textit{Preprint} \eprint{cond-mat/9908003})

\bibitem{ReviewNiu}
Xiao D, Chang M~C and Niu Q 2010 {\em Rev. Mod. Phys.\/} {\bf 82} 1959
  (\textit{Preprint} \eprint{cond-mat/0907.2021})

\bibitem{Chern}
Goryo J and Kohmoto M 2008 {\em Mod. Phys. Lett. B\/} {\bf 22} 303
  (\textit{Preprint} \eprint{cond-mat/0606758})

\bibitem{Thouless}
Thouless D~J 1983 {\em Phys. Rev. B\/} {\bf 27} 6083

\bibitem{TKNN}
Thouless D~J, Kohmoto M, Nightingale M~P and den Nijs M 1982 {\em Phys. Rev.
  Lett.\/} {\bf 49} 405

\bibitem{Kohmoto}
Kohmoto M 1985 {\em Ann. Phys.\/} {\bf 160} 343

\bibitem{Keller}
Keller M~W 2008 {\em Metrologia\/}  102

\bibitem{Joye}
Joye A, Brosco V and Hekking F 2010  (\textit{Preprint}
  \eprint{math-ph/1002.1223})

\bibitem{Hekking}
Brosco V, Fazio R, Hekking F~W~J and Joye A 2008 {\em Phys. Rev. Lett.\/} {\bf
  100} 027002 (\textit{Preprint} \eprint{cond-mat/0702333})

\bibitem{Ground}
Pirkkalainen J~M, Solinas P, Pekola J~P and M\"ott\"onen M 2010 {\em Phys. Rev.
  B\/} {\bf 81} 174506 (\textit{Preprint} \eprint{cond-mat/1002.0957})

\bibitem{Wendin}
Wendin G and Schumeiko V~S 2005  (\textit{Preprint} \eprint{cond-mat/0508729})

\bibitem{Kato}
Kato T 1950 {\em J. Phys. Soc. Japan\/} {\bf 5} 435

\bibitem{Messiah}
Messiah A 1962 {\em Quantum Mechanics\/} vol~2 (Amsterdam: North-Holland)

\bibitem{Wilson}
Wilson K~G 1974 {\em Phys. Rev. D\/} {\bf 10} 2445

\bibitem{Zhou}
Zhou H~Q, Cho S~Y and McKenzie R~H 2003 {\em Phys. Rev. Lett.\/} {\bf 91}
  186803

\bibitem{Pancharatnam}
Pancharatnam S 1956 {\em Proc. Ind. Acad. Sci. A\/} {\bf 44} 247--262

\bibitem{Bargmann}
Bargmann V 1964 {\em J. Math. Phys.\/} {\bf 5} 862

\bibitem{Numerical}
Press W, Flannery B, Teukolsky S and Vetterling W 1993 {\em {Numerical Recipes
  in Fortran 77: The Art of Scientific Computing}\/} vol~1 (Cambridge:
  Cambridge University Press)

\bibitem{Mukunda}
Simon R and Mukunda N 1993 {\em Phys. Rev. Lett.\/} {\bf 70} 880

\bibitem{Wang}
Wang X, Vanderbilt D, Yates J~R and Souza I 2007 {\em Phys. Rev. B\/} {\bf 76}
  195109 (\textit{Preprint} \eprint{cond-mat/0708.0858})

\bibitem{Broda}
Broda B 2002  (\textit{Preprint} \eprint{math-ph/0012035})

\bibitem{Karp}
Karp R~L, Mansouri F and Rno J~S 2000 {\em Turk. Jour. Phys.\/} {\bf 24} 365
  (\textit{Preprint} \eprint{hep-th/9903221})

\end{thebibliography}

\end{document}